# Ellipticity-dependent sequential over-barrier ionization of cold rubidium

Junyang Yuan[1,2,3], Shiwei Liu[4,5], Xincheng Wang[3], Zhenjie Shen[1], Yixuan Ma[1,2,3], Huanyu Ma[1,2,3], Qiuxiang Meng[1,2,3], Tian-Min Yan[1], Yizhu Zhang[1,6], Alexander Dorn[7], Matthias Weidemüller[8,9,10], Difa Ye[4*], Yuhai Jiang[1,2,3,9*]

[1] Shanghai Advanced Research Institute, Chinese Academy of Sciences, Shanghai 201210, China

[2] University of Chinese Academy of Sciences, Beijing 100049, China

[3] School of Physical Science and Technology, ShanghaiTech University, Shanghai 201210, China

[4] Laboratory of Computational Physics, Institute of Applied Physics and Computational Mathematics, Beijing 100088, China

[5] Graduate School, China Academy of Engineering Physics, Beijing 100193, China

[6] Center for Terahertz Waves and College of Precision Instrument and Optoelectronics Engineering, Key Laboratory of Optoelectronics Information and Technical Science, Ministry of Education, Tianjin University, Tianjin 300072, China

[7] Max-Planck-Institut für Kernphysik, Saupfercheckweg 1, DE-69117 Heidelberg, Germany

[8] Hefei National Laboratory for Physical Sciences at the Microscale and Shanghai Branch, University of Science and Technology of China, 201315 Shanghai, China

[9] CAS Center for Excellence and Synergetic Innovation Center in Quantum Information and Quantum Physics, University of Science and Technology of China, 201315 Shanghai, China

[10] Physikalisches Institut, Universität Heidelberg, Im Neuenheimer Feld 226, 69120 Heidelberg, Germany

**Email: jiangyh@sari.ac.cn; yedifa@yahoo.com;

**Abstract** We perform high-resolution measurements of momentum distribution on Rb$^{n+}$ recoil ions up to charge state $n$ = 4, where laser-cooled rubidium atoms are ionized by femtosecond elliptically polarized lasers with the pulse duration of 35 fs and the intensity of 3.3×10$^{15}$ W/cm$^2$ in the over-barrier ionization (OBI) regime. The momentum distributions of the recoil ions are found to exhibit multi-band structures as the ellipticity varies from the linear to circular polarizations.

The origin of these band structures can be explained quantitatively by the classical OBI model and dedicated classical trajectory Monte Carlo simulations with Heisenberg potential. Specifically, with back analysis of the classical trajectories, we reveal the ionization time and the OBI geometry of the sequentially released electrons, disentangling the mechanisms behind the tilted angle of the band structures. These results indicate that the classical treatment can describe the strong-field multiple ionization processes of alkali atoms.

# I. Introduction

Atoms or molecules exposed to intense femtosecond laser fields can be ionized to highly charged states. The ionization processes can be roughly divided into two categories: sequential ionization (SI) and non-sequential ionization (NSI). In the linearly polarized (LP) laser fields with moderate intensity, ionization is dominated by the NSI process [1,2]. When the laser field is close-to-circularly polarized or its intensity is sufficiently high, SI becomes the dominant process [3-5]. In SI, it is assumed that the electrons are ionized one by one. Due to its simplicity, the process of SI has attracted less attention during the past decades. However, SI induced by elliptically polarized (EP) fields can provide more information on the ionization process that is unavailable in NSI by LP fields. For example, by measuring the recoil-ion momentum distributions (RIMDs) from SI in the EP fields, one can retrieve the information about the ionization fields and the ionization times of the emitted electrons [6-9], related partially to the multielectron ionization dynamics in attoclock measurements [10-12].

In recent years, ionization by EP fields has attracted much attention and revealed new phenomena [13-16]. In Ref. [14], Pfeiffer *et al*. have systematically investigated SDI of Ar by EP fields over a wide range of laser intensities and the ionization times of both emitted electrons are extracted as a function of laser intensity. In that experiment, it was shown that the RIMDs of doubly charged ion evolved from a three-band structure to a four-band structure as the laser intensity increased. Moreover, the measured release time of the first electron agrees well with the prediction of the independent electron model, whereas the second electron is much earlier than predicted. In Ref. [17], it has been shown that there is a clear angular correlation between the two electrons from SDI, which implies that the successive ionization steps are not independent in SDI. The classical ensemble model considering electron correlation successfully explained the experimental results [18]. Classical calculations also showed that for multiple ionization the corresponding RIMDs along the minor polarization direction exhibit specific peak structures [7]. In a recent experiment for triple ionization of $Ne^+$, a six-band structure in RIMDs was observed [19], where the saturation intensity for ionization as well as the ionization time for each ionization step were extracted.

So far, the ionization process in strong EP fields has been studied mainly with rare or molecular gases, because these gases could be cooled efficiently by supersonic gas jets [17,20-24] in order to obtain good resolution for the RIMD. In contrast, alkali atoms have rarely been studied under the context of strong field multiple ionization processes, particularly for RIMDs, as the cooling of the solid phase atomic target poses technical challenges. With the help of our newly built magneto-optical trap recoil ion momentum spectroscopy (MOTRIMS) platform combining cold atoms, strong laser pulse, and ultrafast technologies [25], we are able to extend strong-field multiple ionizations and ultrafast processes to alkali atoms. Except for some very specific energy levels of rubidium, the present experiment provides clean spectra due to the employed laser-cooling scheme, benefiting from the absence of other isotopes, as typically present in experiments with supersonic beams of noble gases. The MOTRIMS technique thus represents a powerful tool to study the multiple ionization process of heavy alkali atoms in strong laser field [26-29].

In this article, we present a momentum-resolved study of strong-field multiple ionizations of cold Rb atoms, up to quadruple ionization, at laser intensities in the OBI region. The RIMDs in the polarization plane recorded by MOTRIMS exhibit rich multi-band structures as the ellipticity varies. With the help of theoretical analyses based on the static OBI model as well as dynamical classical trajectory Monte Carlo simulations with Heisenberg potential (CTMC-H), we identify the physical mechanisms responsible for the band structures and tilted angles in the RIMDs. The ionization time and the OBI geometry in sequential multiple ionization (SMI) will be analyzed.

This paper is organized as follows. In Sec. II, the experimental setup and the theoretical methods are described in detail. In Sec. III, we present our main experimental data and simulated results. Finally, we provide a conclusion in Sec. IV. Atomic units (a.u.) are used throughout the paper, unless specified otherwise.

## Ⅱ. METHODOLOGY

### A. Experimental setup

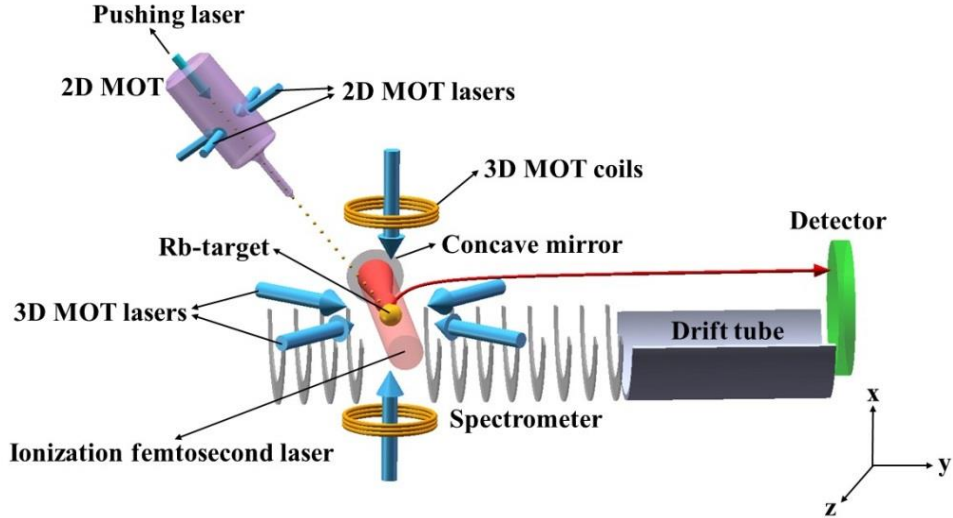

Fig. 1. (Color online). Simplified schematic view of the MOTRIMS experimental setup. Rubidium atoms are pre-cooled within a typical 2D MOT configuration and pushed by the pushing laser into the reaction region, where the atoms are further cooled and trapped in a standard 3D MOT and subsequently ionized by back-reflected femtosecond-laser pulses. The polarization and the propagation directions of the laser are defined as in the $y$ and $z$ directions, respectively. Electric fields (~1 V/cm) guide the recoil-ions to time- and position-sensitive detectors.

Our MOTRIMS platform consists of a femtosecond laser system, a reaction microscope for momentum-resolved ion detection, and a magneto-optical trap for the cold Rb target. A schematic view of the experimental setup is shown in Fig. 1. The vacuum system consists of the pre-cooling chamber and the science chamber partitioned by a valve. Evaporated Rb atoms in a glass cell are pre-cooled and trapped within a typical two-dimensional magneto-optical trap (2D MOT) configuration, and then pushed by a red detuning laser beam into the target region. In the science chamber, 2D cooled Rb can be further cooled and trapped with a standard three-dimensional magneto-optical trap (3D MOT) configuration. By adjusting parameters of the 3D MOT target, molasses and the 2D MOT target, various densities can be selected for the target. The molasses configuration resembles a 2D beam but with cooling lasers turned on. The density of about $10^8$ atoms/cm$^3$ and background vacuum of about $2\times10^{-10}$ mbar are achieved in the science chamber, respectively. The polarized laser beam is focused using a spherical on-axis concave mirror with 75 mm focal length onto the molasses target. Ions produced in the focus of the laser pulse are extracted by a weak uniform electric field of about 1 V/cm onto a multichannel plate detector equipped with delay line anode, which provides both time and position information. The three-dimensional recoil momentum vector of each ion is reconstructed by the information of the corresponding time of flight and the position on the ion detector. For more details of MOTRIMS, see Ref. [25].

A 800 nm Ti: sapphire laser system with 1 kHz repetition rate, 4 mJ pulse energy, and 35 fs pulse duration was employed as an ionization laser source. The combination of a $\lambda/2$ plate and an alpha-BBO Glan-Taylor laser polarizer was used to control the laser intensity. The laser ellipticity was monitored by a zero-order quarter-wave plate. In the experiment we varied the ellipticity at a constant intensity $3.3 \times 10^{15}$ W/cm$^2$, thereby the Keldysh parameters $\gamma$ are ranging from 0.26 to 0.52 for the involved charge states. The laser peak intensity in the interaction region was determined by measuring the "donut"-shape RIMDs of Rb$^{2+}$ with circularly polarized (CP) fields [30]. The uncertainty of the peak intensity was estimated to be ±20%.

## B. Classical trajectory Monte Carlo approach with Heisenberg potential

To understand the multi-electron dynamical process of laser-driven Rb system, we have performed some simulations with classical trajectory Monte Carlo approach with Heisenberg potential (CTMC-H) [31,32]. The total Hamiltonian we exploit can be written as

$$H = \sum_{i=1}^{N} \left( \frac{p_i^2}{2} - \frac{Z}{|r_i|} + \frac{\xi_i^2}{4\alpha r_i^2} e^{\alpha[1-(\frac{|r_i||p_i|}{\xi_i})^4]} \right) + \sum_{i,j=1; i<j}^{N} \frac{1}{|r_i - r_j|} - \sum_{i=1}^{N} E(t) \cdot r_i, \qquad (1)$$

where the position and momentum of electron $i$ are denoted by $\mathbf{r}_i$ and $\mathbf{p}_i$, respectively. The Heisenberg potential $V(\mathbf{r}_i, \mathbf{p}_i, \xi_i) = \dfrac{\xi_i^2}{4\alpha \mathbf{r}_i^2} e^{\alpha[1-(\frac{|\mathbf{r}_i||\mathbf{p}_i|}{\xi_i})^4]}$ is adopted for the electron-nucleus interaction to mimic the Heisenberg uncertainty which prevents the electron from visiting parts of the classical phase space that would be forbidden in quantum mechanics [33-35]. Here, $\alpha$ controls the rigidity of the potential and $\xi_i (i=1\cdots\cdots N)$ are chosen to fit the ionization potentials of Rb atom.

The electric field of the laser is given by

$$E(t) = \frac{E_0 f(t)}{\sqrt{1+\varepsilon^2}} [\varepsilon \sin(\omega t)\hat{x} - \cos(\omega t)\hat{y}], \qquad (2)$$

where $E_0$, $\omega$, and $\varepsilon$ are the amplitude, frequency, and ellipticity of the laser field, respectively. The envelope function is $f(t) = \sin^2(\dfrac{\pi t}{T})$ and $T$ is the pulse length, which is set as twenty optical cycles in our following calculations. Since the laser intensity considered is not enough to ionize the 4$s$ and other inner shell electrons of Rb, we restrict our simulations to seven active electrons and thus set $N = Z = 7$. By adding the electrons one by one and minimizing the value of the Hamiltonian at each step [34-36], the atomic parameters are determined as $\xi_1 = 2.5053$, $\xi_2 = 2.4784$, $\xi_3 = 2.4413$, $\xi_4 = 2.4216$, $\xi_5 = 2.3205$, $\xi_6 = 2.2707$ and $\xi_7 = 2.4520$. The corresponding electron distances to the nucleus are $r_1 = 5.77$, $r_2 = 1.43$, $r_3 = 1.29$, $r_4 = 1.24$, $r_5 = 1.02$, $r_6 = 0.94$ and $r_7 = 1.33$, which highly resembles the shell structure of a Rb atom.

We then further exploit the Monte Carlo sampling technique of the random Euler's rotations for the above configuration [37], in order to get an ensemble of electrons with different initial positions and momenta. The dynamics of the system is governed by the following canonical equations:

$$\frac{d\mathbf{r}_i}{dt} = \frac{\partial H}{\partial \mathbf{p}_i}, \quad \frac{d\mathbf{p}_i}{dt} = -\frac{\partial H}{\partial \mathbf{r}_i}. \qquad (3)$$

We solve these equations numerically by employing the standard fourth-fifth Runge-Kutta algorithm and the ionization events are identified by the final energy of electrons. More than $10^6$ classical trajectories are traced to ensure the convergence of the statistical results.

## III. RESULTS AND DISCUSSIONS

### A. Experimental observation of RIMDs

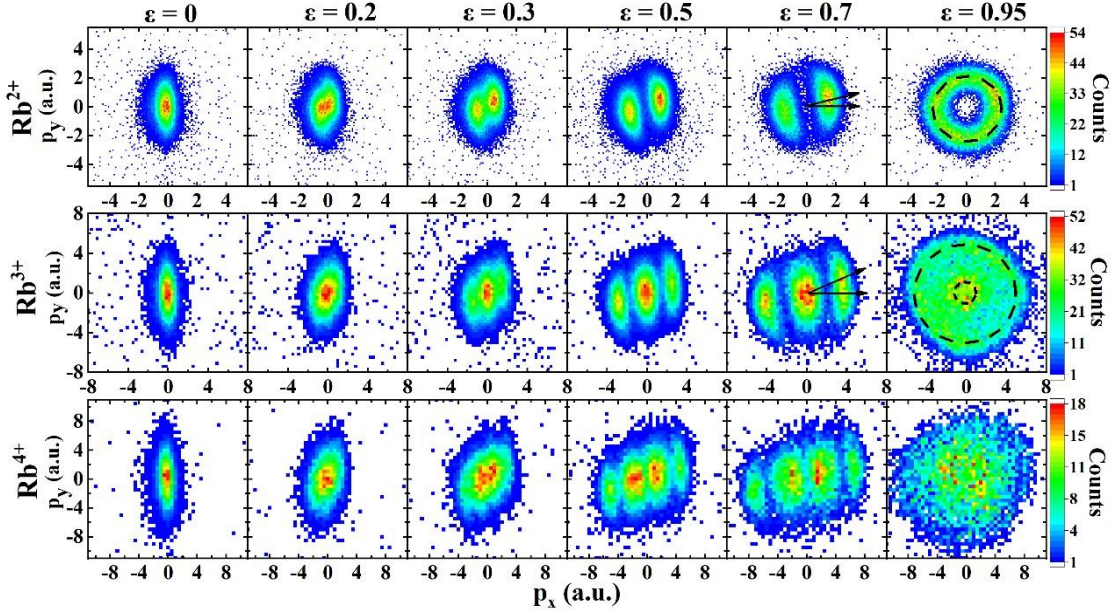

Fig. 2 (Color online). Measured RIMDs in the polarization plane for different ellipticities $\varepsilon$ (columns) from linearly polarized to almost circularly polarized fields for double, triple, and quadruple ionizations (rows) of neutral Rb atoms. The major/minor polarization direction is along the $y/x$ axis. The arrows indicate the tilted angles and dashed circles represent predicted momenta based on the OBI calculations.

Fig. 2 displays the measured $\varepsilon$-dependent RIMDs of $Rb^{2+}$, $Rb^{3+}$ and $Rb^{4+}$ ions in the polarization $x$-$y$ plane, where $p_x$ and $p_y$ are the momenta along the minor and major axes of the laser polarization ellipse, respectively. For $Rb^{2+}$, $Rb^{3+}$ and $Rb^{4+}$ ions at the ellipticity $\varepsilon = 0$, the RIMDs display a two-dimensional Gaussian distribution with its maximum at zero momentum. As expected, the distribution is more expanded in the polarization direction ($y$ axis). For increasing ellipticities, the RIMDs split from one band into multi-band structures along the minor axis ($x$ axis) of the polarization ellipse and the band structures become obvious at higher ellipticities, whereas along the major axis (*y* axis) each band is still close to Gaussian. This is due to the fact that ionization occurs preferentially when the electric field vector points along the major polarization axis (*y* axis), producing electrons with a momentum pointing in the *x* direction after acceleration by the electric field of the laser pulse [7,8,13,38]. For $\varepsilon >$ 0.3, a two-, three- and four-band structures are presented for $Rb^{2+}$, $Rb^{3+}$ and $Rb^{4+}$ ions, respectively. Moreover, there is always a tilted angle with respect to the minor polarization axis, e.g., at $\varepsilon = 0.7$, tilted angles are in the region of 10° - 20° for $Rb^{2+}$ and 25° - 35° for $Rb^{3+}$, as indicated by arrows in Fig. 2. Our CTMC-H simulations indicate that the scattering of the parent ion contributes to tilted angles, details will be discussed below. As ellipticity increases to $\varepsilon = 0.95$ (close-to CP), the measured RIMDs become circularly symmetric. A donut structure and a two-concentric-ring structure are respectively observed for the RIMDs of $Rb^{2+}$ and $Rb^{3+}$ shown with dashed lines in Fig. 2, whereas the result of $Rb^{4+}$ seems Gaussian-like and contains vague structure due to the limited experimental resolution.

## B. OBI model explanation of the band structure

The origin of the band structure can be understood quantitatively by using the classical OBI model. For sequential multiple ionization, the momentum of the recoil ion can be considered to be the sum of the momenta gained from each ionization steps of electrons. Therefore, the number of observable peaks results from the various momentum combinations of the emitted electrons if experimental resolution is sufficient. For elliptical polarization, the peak field strength *E* along the major polarization axis (*y* axis) is stronger than that along the minor polarization axis (*x* axis). In this case, ionization happens most probably around field maxima in the *y* direction because the ionization probability can be considered to depend exponentially on field strength. Therefore $E_y$ can be regarded as the instantaneous laser field at the time of ionization. Classical calculations show that the momentum (along the minor axis) of each electron at the end of the pulse can be expressed as $p_x = \varepsilon E_y(t)/\omega$, where $\varepsilon$, $\omega$, and $E_y(t)$ are the field ellipticity, the angular frequency, and the instantaneous field strength (along the major axis) at the time of ionization [7], respectively.

For Rb atoms, the first, second, third and fourth ionization potentials are 2.6 eV (excited state), 27.3 eV, 39.2 eV and 52.2 eV, respectively. The corresponding OBI field strength $E_{OBI}$ ($E_{OBI} = I_p^2/4Z$, where $I_p$ is the ionization energy for producing an ion of charge $Z$ [6,39].) are 0.0023a.u., 0.1259 a.u., 0.1731 a.u. and 0.2302 a.u. respectively. For the laser intensity of $3.3 \pm 0.66 \times 10^{15}$ W/cm$^2$ used in our experiments, the corresponding peak electric field strengths are about $0.3066 \pm 0.0613$ a.u. and $0.2223 \pm 0.0445$ a.u. for the LP ($\varepsilon = 0$) and close-to CP ($\varepsilon = 0.95$) fields, which are higher than the $E_{OBI}$ of these four electrons evaluated above, respectively. Thus, Rb$^+$, Rb$^{2+}$, Rb$^{3+}$ ions are created within the OBI regime, while and Rb$^{4+}$ occurs at the edge of the OBI regime. In the OBI region, the band-structures in the RIMDs are no longer intensity dependent [40]. In the case of close-to CP fields, the RIMD of Rb$^{2+}$ show a circularly symmetric donut structure and the radius of the donut is about 2.2 a.u. according to the OBI model. Similarly, the RIMD of Rb$^{3+}$ show a two-concentric-rings structure and the corresponding radius of the inner ring and the outer ring are 0.8 a.u. and 5.2 a.u. respectively. The black dashed curves at $\varepsilon = 0.95$ in Fig. 2 indicate the momentum values calculated for the classical OBI model, consistent with present measurements.

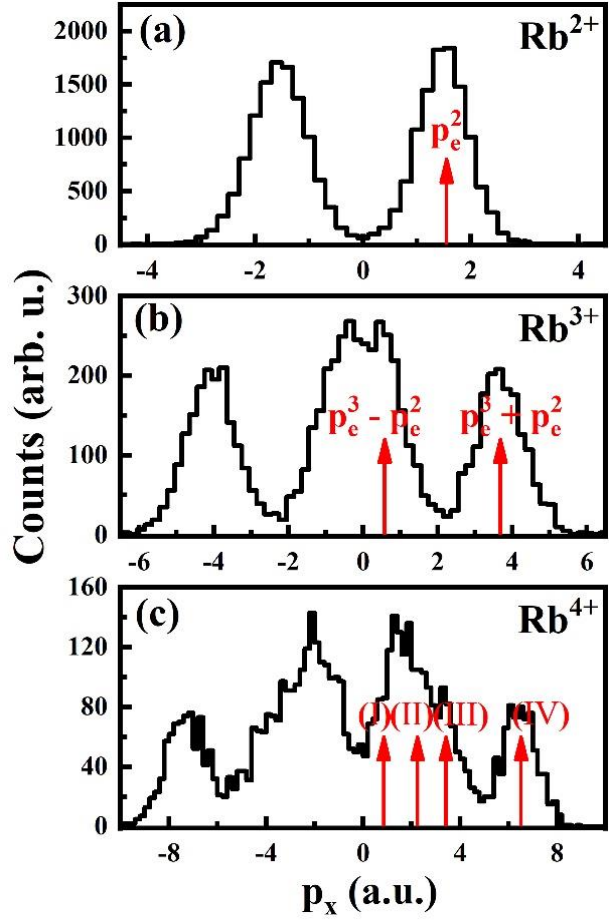

Fig. 3. (Color online). One-dimensional momentum spectra of $Rb^{2+}$ [panel (a)], $Rb^{3+}$ [panel (b)] and $Rb^{4+}$ [panel (c)] ions, projected onto the *x* axis (along the minor axis) at ellipticity ε = 0.7 in Fig. 2. The vertical arrows indicate the peak locations expected for the classical OBI model, as discussed in the text. The four peaks in panel (c), from left to right, corresponding to $p_i^{4+}(1) = (-p_e^2 - p_e^3 + p_e^4)$ (peak I), $p_i^{4+}(2) = (p_e^2 - p_e^3 + p_e^4)$ (peak II), $p_i^{4+}(3) = (-p_e^2 + p_e^3 + p_e^4)$ (peak III), and $p_i^{4+}(4) = (p_e^2 + p_e^3 + p_e^4)$ (peak IV), respectively. $p_i^{n+}$ and $p_e^m$ are defined as the momenta of recoil ion in charge state *n*+ and the *m*th ionizing electron, respectively.

Taking *ε* = 0.7 as an example, we give more quantitative explanations to the origin of the band structures along the minor elliptical axis (*x* axis) with the classical OBI model [6]. Here, multi-band structures at *ε* = 0.7, shown in Fig. 2, are more pronounced. In Fig. 3, we depict the $p_x$ spectra, with momentum projections in the direction of the minor elliptical axis of $Rb^{2+}$, $Rb^{3+}$ and $Rb^{4+}$ ions in Fig. 2. For the laser intensity of 3.3×10$^{15}$ W/cm² used, the corresponding maximum electric field strength along the major axis of the polarization ellipse is about 0.25 ± 0.05 a.u. at *ε* =

0.7. In this case, $Rb^{2+}$, $Rb^{3+}$ and $Rb^{4+}$ ions are assumed to be generated in the OBI region. On the other hand, for the laser intensity and the ellipticity discussed here, the field along the *x* direction drives the emitted electrons transversely and effectively eliminates the possibility of a recollision so that the creation of each charge state can be regarded as determined by the SI process [7]. Consequently, $E_{OBI}$ can be regarded as the instantaneous laser field at the time of ionization. Consequently, the momentum (along *x* axis) of each electron at the end of the pulse will be given by $p = \varepsilon E_{OBI}/\omega$, which results in the corresponding momenta (along *x* axis) of the first, second, third, and fourth ionized electrons at the end of the pulse being $p_e^1 = 0.03$ a.u., $p_e^2 = 1.55$ a.u., $p_e^3 = 2.13$ a.u., and $p_e^4 = 2.83$ a.u., respectively.

According to these electron momenta $p_e^m$ calculated from the classical OBI model, we discuss the RIMDs projected into the positive *x* direction. In Ref. [40], it has been demonstrated that the momentum of the first ionized electron is negligibly small and thus can be ignored when analyzing the ionization mechanism of highly charged states. As a result, momentum conservation of the RIMD of $Rb^{2+}$ leads to $\boldsymbol{p_i^{2+}}$ = - $\boldsymbol{p_e^2}$. Therefore the RIMD of $Rb^{2+}$ is expected to have a single-peak positioned at $p_i^{2+} = 1.55$ a.u. in the minor polarization direction according to the OBI calculation. As shown in Fig. 5, one can note that the measured results of $Rb^{2+}$ displays a clear peak, which is consistent with the classical estimations. Here the recoil momentum from the first electron might result in the broadening of $\boldsymbol{p_i^{2+}}$ although its momentum is negligible. Similarly, the RIMD of $Rb^{3+}$ is obtained by the sum of the second and the third ionized electron momentum vectors. As illustrated in Fig. 3(b), we observe that the measured results of $Rb^{3+}$ show two peaks positioned at $p_i^{3+}(1) = p_e^3 + p_e^2 = 3.68$ a.u. (parallel electron emission) and at $p_i^{3+}(2) = p_e^3 - p_e^2 = 0.58$ a.u. (anti-parallel electron emission), respectively, which are also supported well by the classical OBI model indicated by arrows in Fig. 3(b). The same strategy can be applied to quadruple ionization and calculations predict four combinations: $p_i^{4+}(1) = -p_e^2 - p_e^3 + p_e^4 = 0.85$ a.u., $p_i^{4+}(2) = p_e^2 - p_e^3 + p_e^4 = 2.25$ a.u., $p_i^{4+}(3) = -p_e^2 + p_e^3 + p_e^4 = 3.41$ a.u., and $p_i^{4+}(4) = p_e^2 + p_e^3 + p_e^4 = 6.51$ a.u.. The expected locations are indicated by the vertical arrows in Fig. 3(c). The peak $p_i^{4+}(4)$ is clearly observed, corresponding to the ionization combination of the second, the third, and the fourth electrons emitted sequentially into the same direction. However, the peaks of $p_i^{4+}(1)$, $p_i^{4+}(2)$, and $p_i^{4+}(3)$ are so close to each other that they cannot be resolved due to the finite count rates and the limited experimental resolution.

In general, the structures in RIMD are well described via the OBI model, where all electrons are assumed to be emitted sequentially and independently. Meanwhile, in the case of close-to CP, one also notes that the various emission combinations of electrons contribute almost equally taking their amplitudes into account. So we can conclude that correlations among ionized electrons appear to be not important for the processes studied here.

### C. CTMC-H simulation and the complex ionization dynamics

In order to gain a deeper insight into the underlying physics from the measured RIMDs, we have performed simulations with the CTMC-H model at the experimental parameters. The calculated RIMDs are shown in Fig. 4. For comparison, we present the simulated and the measured results in the same scales for the same charge state. Whereas the RIMD of $Rb^{4+}$ seems consistent with the experiment, the simulated results for $Rb^{2+}$ and $Rb^{3+}$ are broader and exhibit richer structures compared to the experiment, showing four- and six-band structures, respectively. More specifically for $Rb^{2+}$ in close-to-CP fields, the simulated results show a two-concentric-ring structure, in contrast to the single-ring structure observed in experiment.

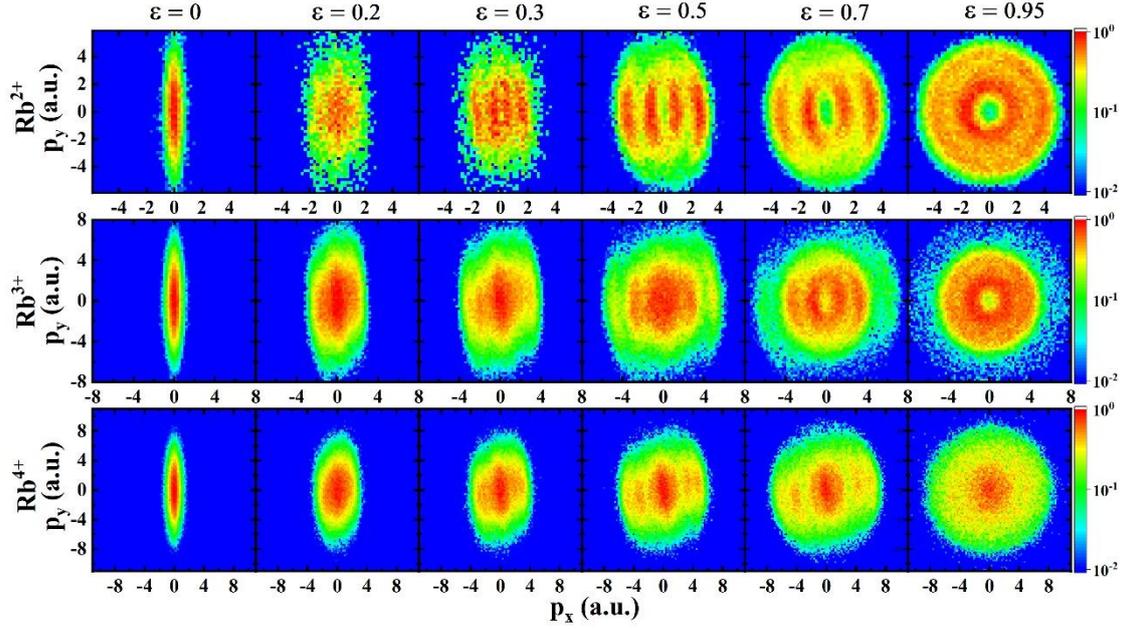

Fig. 4 (Color online). The simulated ellipticity-dependent RIMDs of Rb for double, triple, and quadruple ionizations at the intensity of $3.3 \times 10^{15}$ W/cm$^2$.

The question arises what leads to the deviation between the experimental and the simulated results. In fact, the following analysis shows that the deviation is a result of the volume effect [41]. In Ref. [8,14], it has been shown that the RIMD along the minor elliptical axis exhibits a characteristic dependence on laser intensity: For the 33-fs pulses, there is a bifurcation from a three-peak structure to a four-peak structure as the laser intensity increases. This indicates that the momentum spectrum depends on the laser intensity and the volume effect should be taken into account. However, we note that in our following calculations we do not quantitatively consider the focal volume effect which would require the knowledge on the precise geometry of the laser focus and the target beam.

Instead, we choose a lower laser intensity and demonstrate that the RIMDs strongly depend on the laser intensity. Using the calculated OBI laser intensity of Rb$^{2+}$, i.e., $5\times10^{14}$ W/cm$^2$, we have also simulated the ellipticity-dependent RIMDs of Rb$^{2+}$ and Rb$^{3+}$ ions based on the CTMC-H model, as seen in Fig. 5. For Rb$^{2+}$ and Rb$^{3+}$ ions at $\varepsilon = 0$, one can see both the experimental and the simulated results display a two-dimensional Gaussian distribution. For $\varepsilon > 0.3$, compared with Fig. 4 ($3.3 \times 10^{15}$ W/cm$^2$), the simulated results at $5\times10^{14}$ W/cm$^2$ are more consistent with the experimental results. Though the two results are not identical, the same trends of experiments and theory (e.g., the number of observable peaks and the tilted angle shown in Fig. 2 and Fig. 5) on $Rb^{2+}$ and $Rb^{3+}$ indicate that the CTMC-H simulations allow one to qualitatively capture the ionization mechanisms observed.

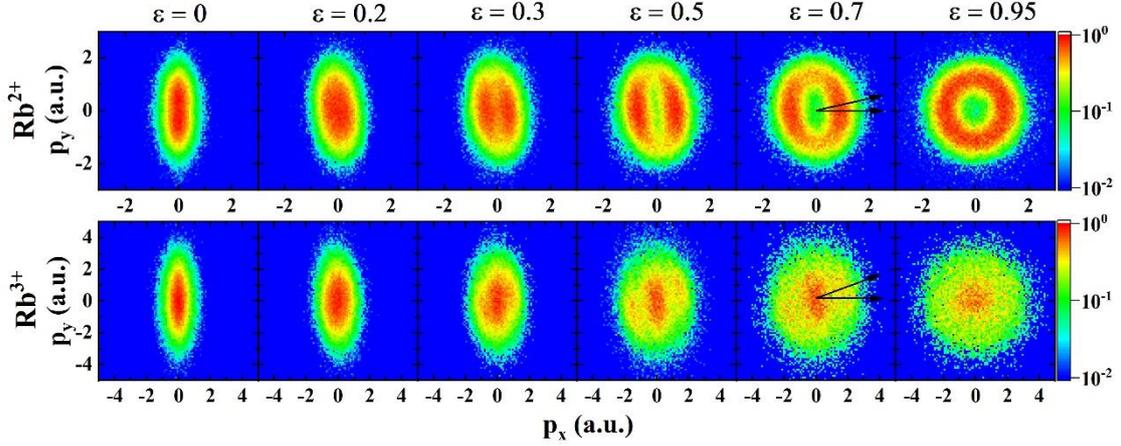

Fig. 5 (Color online). The same as Fig. 4 but at $5.0 \times 10^{14}$ W/cm$^2$. The quadruple ionization events are rare at this intensity and thus are not shown.

With the CTMC-H model justified, we apply it to study the ionization time and the ionization exit which provide knowledge on the complex ionization dynamics not accessible by the OBI model.

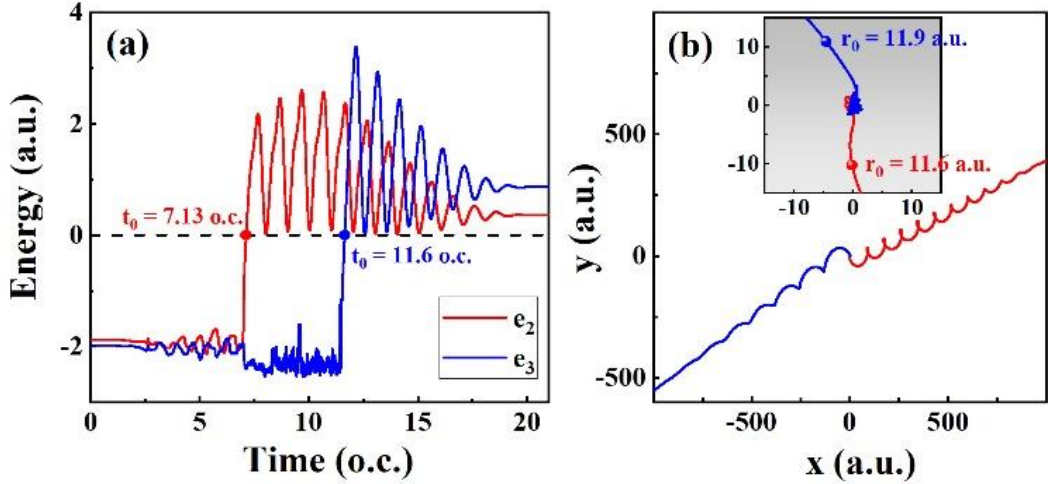

Fig. 6 (Color online). Typical trajectory of a sequential ionization by elliptically polarized strong laser field at $5.0 \times 10^{14}$ W/cm$^2$ and $\varepsilon = 0.7$. The ionization time, defined as the instant when the energy of the electron becomes positive, is marked by the solid circles in (a) and the corresponding ionization geometries are shown in (b).

A typical triple ionization trajectory is depicted in Fig. 6, where only the second and the third ionized electrons are shown. The outermost electron can be safely neglected because it has a very low ionization potential and is always liberated at the early stage of the laser pulse. The energy evolutions of both electrons are shown in Fig. 6(a), where the ionization time $t_0$ is defined at the instant when the electron energy becomes positive. At the vicinity of the ionization time, the electron is released across the saddle point formed by the laser field suppressed Coulomb barrier, initially upward or downward along the major ($y$) axis. During its way away from the barrier, however, the trajectory is distorted counter-clockwise due to the scattering of the parent ion, similar to that explained in the Rutherford-Keldysh model [42]. Finally, the two ionized electrons drift out back-to-back essentially along the direction perpendicular to the initial position vector at the ionization exit, due to the $\pi/2$-phase lag between the laser electric field and the vector potential. More precisely, the emission direction points towards the left lower and right upper corners, which explains the tilted angle in the RIMDs. However, since we focus on the over-barrier ionization here, the tilted angle is not related to the tunneling time or the non-adiabaticity (i.e., the initial momentum of the tunneled electrons) as routinely invoked in the literatures [43-49].

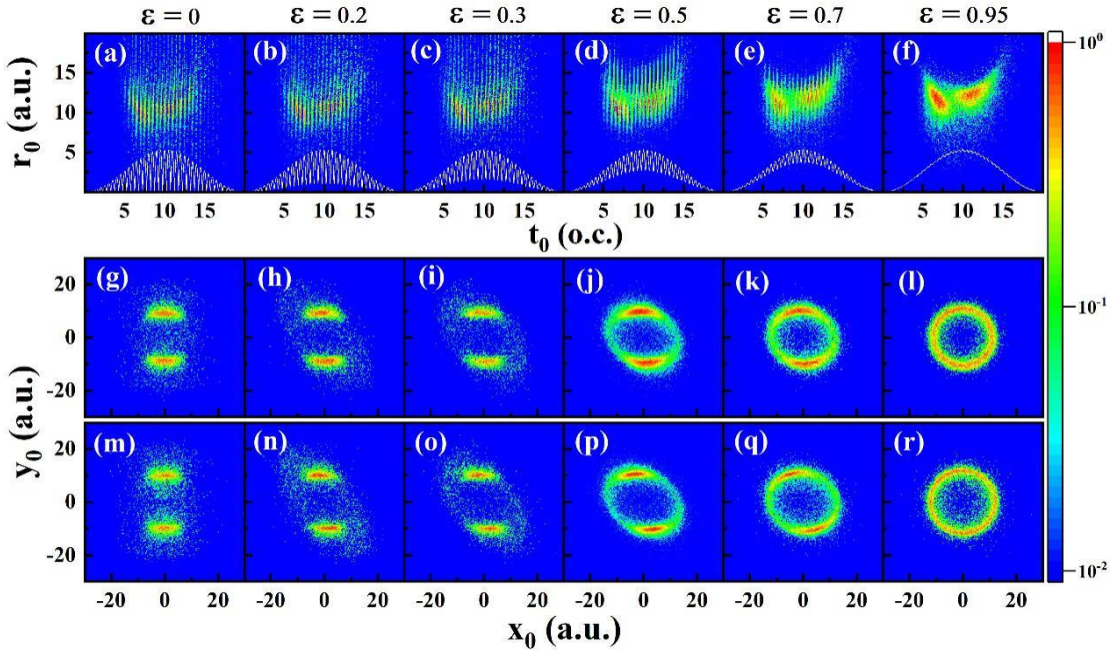

Fig. 7 (Color online). The joint distributions of the ionization time and the electron distance to the nucleus (upper row). The white curves illustrate the waveform of the electric field |E(t)| which explains the sub-cycle stripe structures at lower ellipticities. The exit geometries in the polarization plane are shown in the middle (second ionization) and the lower (third ionization) rows, respectively. The laser parameters are the same as that in Fig. 6.

A more intuitive picture can be obtained with performing statistics on an ensemble of electrons as given in Fig. 7, where the first row shows the joint distributions of the ionization time and the electron distance to the nucleus. The two bright [colored in red, see, e.g., Fig. 7(f)] spots represent the second (left spot) and the third (right spot) ionized electrons, respectively. The results indicate that the third electron is most probably ionized at the peak of the laser pulse while the second electron escapes slightly earlier. The electron distance to the nucleus at the ionization time roughly peaks around 10 a.u.. Interestingly, close inspection on the exit geometry in the polarization plane reveals that at $\varepsilon = 0$ the electrons are mainly released along the major axis as expected, while with the increase of the ellipticity, the released electrons peak at a certain angle with respect to the major axis. The tilted angle of the third ionized electron is larger than that of the second ionized electron [e.g., comparing Fig. 7(q) with Fig. 7(k)] due to the stronger Coulomb scattering. This insight finally disentangles the mechanism leading to the large tilted angle (typically 30°) as observed RIMDs (e.g. revisiting the tilted angle in Fig. 5).

## IV. SUMMARY

In conclusion, employing a recently developed MOTRIMS setup combining cold atoms, strong laser pulses, and ultrafast technologies, we have measured the ellipticity-dependent RIMDs of Rb atoms ionized by strong EP fields. With increasing the ellipticity of the laser pulse, the RIMDs are shown to contain rich ionization information and exhibit specific band structures: a two-, three-, and four-band structure for the doubly, triply, and quadruply charged ions, respectively. We show that these experimentally observed multi-band structures and the tilted angles of the bands with respect to the polarization axis can be well explained by the OBI model and CTMC-H simulations. With back analysis of the classical trajectories, we reveal their relationship with the ionization time and the OBI geometry of the sequentially released electrons. The qualitative agreement between our numerical results and experimental data indicates that a classical treatment remains reasonable in describing the strong-field multiple ionization where the fully quantum approach is of great challenge. Our work might inspire further investigations on the complex multi-electron dynamics in strong-field processes.

## ACKNOWLEDGMENTS

This work was supported by the National Natural Science Foundation of China (NSFC) (11827806, 11874368, 61675213, 11822401, 11674034). We acknowledge the support from Shanghai-XFEL beamline project (SBP) and Shanghai High repetition rate XFEL and Extreme light facility (SHINE).